\documentclass[prl,twocolumn,preprintnumbers,amsmath,amssymb,showpacs,nofootinbib,floatfix]{revtex4}

\usepackage{graphicx,bm}

\makeatletter
\def\graphicscale{\twocolumn@sw{0.3}{0.4}}
\def\graphicthreescale{\twocolumn@sw{0.3}{0.4}}

\begin{document}

\title{
Dynamic off-equilibrium transition in systems slowly driven\\
across thermal first-order phase transitions}

\author{Andrea Pelissetto$^1$ and Ettore Vicari$^2$} 

\address{$^1$ Dipartimento di Fisica di Sapienza, Universit\`a di Roma
        and INFN, Sezione di Roma I, I-00185 Roma, Italy}
\address{$^2$ Dipartimento di Fisica dell'Universit\`a di Pisa
        and INFN, Largo Pontecorvo 3, I-56127 Pisa, Italy}

\date{\today}

\begin{abstract}

We study the off-equilibrium behavior of systems with short-range
interactions, slowly driven across a thermal first-order transition,
where the equilibrium dynamics is exponentially slow.  We consider a
dynamics that starts in the high-$T$ phase at time $t = t_i<0$ and
ends at $t=t_f>0$ in the low-$T$ phase, with a time-dependent
temperature $T(t)/T_c \approx 1 - t/t_s$, where $t_s$ is the protocol
time scale.  A general off-equilibrium scaling (OS) behavior emerges
in the limit of large $t_s$.  We check it at the first-order
transition of the two-dimensional $q$-state Potts model with $q=20$
and 10.  The numerical results show evidence of a dynamic transition,
where the OS functions show a spinodal-like singularity.  Therefore,
the general mean-field picture valid for systems with long-range
interactions is qualitatively recovered, provided the time dependence
is appropriately (logarithmically) rescaled.

\end{abstract}


\maketitle



The dynamical behavior of statistical systems driven across phase
transitions is a typical off-equilibrium phenomenon.  Indeed, the
large-scale modes present at the transition are unable to reach
equilibrium as the system changes phase, even when the time scale
$t_s$ of the variation of the system parameters is very large.  Such
phenomena are of great interest in many different physical
contexts~\cite{Kibble-76,Zurek-85,Binder-87,PSSV-11,Biroli-15,
  CDTY-91,BCSS-94,Bray-94,BBFGP-96,Ruutu-etal-96,
  CPK-00,CGMB-01,MMR-02,
  MPK-03,CG-04,CGM-06,MMARK-06,SHLVS-06,WNSBDA-08,GPK-10,CWCD-11,
  Chae-etal-12,MBMG-13,EH-13,Ulm-etal-13,Pyka-etal-13,LDSDF-13,
  Corman-etal-14,NGSH-15,Braun-etal-15,DWGGP-16}: One observes
hysteresis and coarsening phenomena, the Kibble-Zurek defect
production, etc.  At continuous transitions, thermodynamic quantities
obey general off-equilibrium scaling laws as a function of $t_s$,
controlled by the universal static and dynamic exponents of the
equilibrium transition~\cite{GZHF-10,CEGS-12}.  Similar results hold
along the {\em magnetic} first-order transition line of systems with
continuous O($N$) symmetries ($N>1$)~\cite{PV-16}.

This Letter considers systems with short-range interactions undergoing
a {\em thermal} first-order transition (FOT) driven by the temperature
$T$. At the FOT temperature $T_c$, the energy density is discontinuous
and any local dynamics is very slow, due to an exponentially large
tunneling time between the two phases: $\tau(L) \sim \exp(\sigma
L^{d-1})$ for a system of size $L^d$, where the constant $\sigma$ is
related to the interface free energy.  We study the off-equilibrium
behavior arising when $T$ is slowly varied across $T_c\equiv
\beta_c^{-1}$. We consider a linear time dependence
\begin{equation}
\delta(t)\equiv \beta(t)/\beta_c - 1  = t/t_s,\qquad \beta\equiv 1/T,
\label{deltat}
\end{equation} 
starting the dynamics at a time $t_i<0$ in the high-$T$ phase and
ending it at $t_f > 0$ in the low-$T$ phase. $t_s$ is the time scale
of the temperature variation. This protocol is general since a generic
time dependence can be approximated by a linear function around $T_c$.

In the mean-field approximation, which becomes exact for long-range
interactions~\cite{Binder-87}, after crossing $T_c$ the system
persists in a metastable state with an infinite mean life, up to a
spinodal-like point $T_{\rm sp}<T_c$, thus up to a time $t>0$ such
that $\delta(t)=T_c/T_{\rm sp}-1$, where a rapid transition to the
low-$T$ phase occurs.  This picture requires a substantial revision in
the case of short-range interactions, because metastable states may
decay when $T(t)<T_c$, due to droplet formation~\cite{Binder-87}.

We show that short-ranged systems at a thermal FOT show an
off-equilibrium scaling (OS) behavior, which significantly differs
from that obtained in the mean-field approximation. For finite $t_s$
we observe a sharp transition to the low-temperature phase at a
temperature $T(t_s) < T_c$, but the temperature $T(t_s)$ approaches
(logarithmically) $T_c$ as $t_s$ becomes large.  Moreover, the time
dependence of the OS functions develop a singular behavior
characterized by peculiar scaling properties.

To test the general OS ideas, we consider the 2D Potts model, which is
an ideal theoretical laboratory to study thermal FOTs. Its Hamiltonian
reads
\begin{equation}
H =  - \sum_{\langle {\bm x}{\bm y}\rangle} \delta(s_{{\bm x}}, s_{ {\bm y}}), 
\label{potts}
\end{equation}
where the sum is over the nearest-neighbor sites of a square lattice,
$s_{\bm x}$ ({\em color}) are integer variables $1\le s_{{\bm x}} \le
q$, $\delta(a,b)=1$ if $a=b$ and zero otherwise.  It undergoes a phase
transition~\cite{Baxter-book,Wu-82} at $\beta_c = \ln(1+\sqrt{q})$,
between a disordered phase and an ordered phase with $q$ equivalent
{\em vacua}. The transition is of first order for $q>4$.  We consider
$L\times L$ square lattices with periodic boundary conditions (PBC),
which preserve the $q$-permutation symmetry.  In infinite volume the
energy density $E = \langle H \rangle/L^2$ is discontinuous at $T_c$,
with different~\cite{results-q20} $E_c^\pm \equiv E(T_c^\pm)$. We
define the {\em renormalized} energy density
\begin{eqnarray}
E_r \equiv  
\Delta_e^{-1}\,(E -E_c^-),\qquad \Delta_e \equiv E_c^+ -
E_c^-,
\label{ener}
\end{eqnarray}
which satisfies $E_r=0,1$ for $T\to T_c^-$ and $T\to T_c^+$,
respectively.  The magnetization
\begin{eqnarray}
M_k = {1\over L^2} \langle \sum_{\bm x} \mu_k({\bm x}) \rangle,
\qquad \mu_k({\bm x}) \equiv {q \delta(s_{\bm x},k) - 1\over q-1},
\label{mkdef}
\end{eqnarray}
vanishes due to the $q$-state permutation symmetry, for any $T$.  We
consider the correlation function $G_{kp}({\bm x},{\bm y}) \equiv
\langle \mu_k({\bm x}) \mu_p({\bm y}) \rangle$, and in particular its
space integral
\begin{equation}
I_G = L^{-2}  \sum_{k=1}^q  \sum_{{\bm x},{\bm y}} 
G_{kk}({\bm x},{\bm y}).
\label{Igdef}
\end{equation}
Equilibrium finite-size scaling (EFSS) holds also at
FOTs~\cite{NN-75,FB-82,PF-83,FP-85,CLB-86,BK-90,CNPV-14}.  For
cubic-like lattices, the relevant scaling variable is $r_1 = L^d
\delta$, where $\delta \equiv \beta/\beta_c-1$. The energy density and
$I_G$ scale correspondingly as
\begin{eqnarray}
E_r(T,L) \approx {\cal E}_{\rm eq}(r_1), \quad
I_G(T,L) \approx L^d {\cal C}_{\rm eq}(r_1),
\label{fss}
\end{eqnarray}
in the EFSS limit $L\to\infty$ keeping $r_1$ fixed.~\cite{footnoteefss}

The system is driven across the transition by the temperature protocol
(\ref{deltat}), starting from equilibrated configurations at $\beta =
\beta_i=\beta(t_i)<\beta_c$.  Observables, such as $E_r$ and $I_G$,
are averaged at fixed $t$ over the starting configurations.  We
anticipate that the OS behavior across the FOT does not depend on the
value of $\beta_i<\beta_c$.

To specify the OS laws that describe the dynamic behavior for
$\beta(t)\approx \beta_c$, we must identify the correct scaling
variables.  First, we use the variable $r_1$, parametrizing the EFSS
functions, as equilibrium should be recovered in the appropriate
limit. To define a second scaling variable, we should identify the
appropriate time scale.  When the global symmetry is preserved by the
boundary conditions or in the absence of boundaries such as PBC, the
the slowest mode in the system is the tunneling between the two
phases.  This is expected to proceed via mixed-phase strip-like
configurations with two interfaces, whose probability is suppressed by
a factor $\exp(-\sigma L)$,~\cite{BN-92,footnotedbc} where $\sigma = 2
\beta_c \kappa$ and $\kappa$ is the interface tension (which is
exactly known for 2D Potts models~\cite{results-q20}). Thus the
relevant time is $\tau(L) = L^\alpha \exp(\sigma L)$ where $\alpha$ is
an appropriate exponent.  Therefore, the OS behavior is expected to be
controlled by the scaling variables
\begin{eqnarray}
r_1 =(t/t_s)L^{2},\qquad  r_2 = t/\tau(L),
\label{s12}
\end{eqnarray}
where $t_s$ is the time scale of the protocol (\ref{deltat}).  The
deviations from equilibrium are conveniently controlled by 
\begin{eqnarray}
s_1 = {r_2/r_1} = t_s/ [L^2\tau(L)] .
\label{uwdef}
\end{eqnarray}
We expect $E_r(t,t_s,L)$ and
$I_G(t,t_s,L)$, defined as in Eqs.~(\ref{ener}) and (\ref{Igdef})
and averaged at fixed $t$, to
scale as
\begin{equation}
E_r \approx {\cal E}_s(s_1,r_1), \qquad
I_G \approx L^2 {\cal C}_s(s_1,r_1),
\label{Ofss}
\end{equation}
in the OS limit $t,t_s,L\to \infty$ keeping $r_1$ and $s_1$ fixed,
thereby extending the EFSS relations (\ref{fss}). EFSS should be
recovered for $s_1 \to \infty$, where ${\cal E}_s(s_1,r_1)$ and ${\cal
  C}_s(s_1,r_1)$ converge to their equilibrium counterparts ${\cal
  E}_{\rm eq}(r_1)$ and ${\cal C}_{\rm eq}(r_1)$.
These OS arguments are quite general and 
can be extended to any thermal FOT, in any 
dimension~\cite{footnotedim}. 

The above OS theory is checked by a numerical analysis of Monte Carlo
(MC) simulations of the 2D Potts model (\ref{potts}) for $q=20$ and
$q=10$. We mostly present results for $q=20$.  The dynamics is
provided by the heat-bath algorithm~\cite{footnoteheatbath}, which is
a representative of a purely relaxational dynamics.  The time unit is
a sweep of the whole lattice.  The temperature is changed according to
Eq.~(\ref{deltat}) every sweep, incrementing $t$ by one.

\begin{figure}[tbp]
\includegraphics*[scale=\graphicscale]{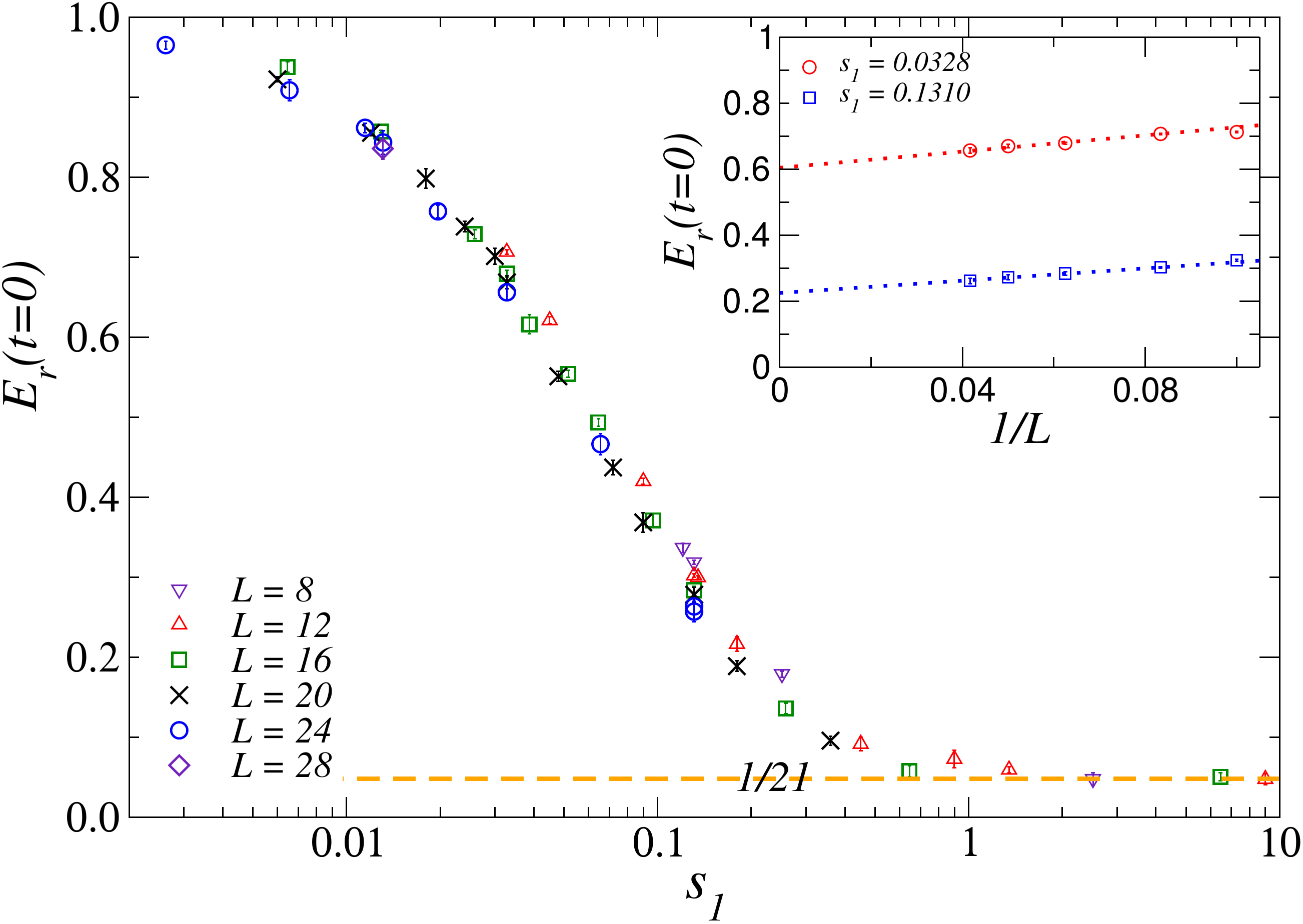}
\caption{(Color online) MC data of $E_r$ for $q=20$ at $t=0$ versus
  $s_1=t_s/(L^{2+\alpha} e^{\sigma L})$, using the optimal value
  $\alpha=2$~\cite{footnotepre}.  For $s_1 \to \infty$, the data
  converge to ${\cal E}_{\rm eq}(0)=1/(1+q)$ (dashed line), since
  equilibrium is approached for $t_s\gg\tau(L)$.  The inset shows the
  approach to the large-$L$ limit at fixed $s_1$.  }
\label{gu4l}
\end{figure}

We first consider data at $t=0$, i.e., $r_1 = 0$, as a function of
$s_1$, see Fig.~\ref{gu4l}.  Their optimal scaling is obtained when
the power of the prefactor of $\tau(L)$ is $\alpha \approx
2$~\cite{footnotebias}.  We also verify the OS of $E_r$ and $I_G$ (and
other observables) with respect to $r_1$, cf. Eq.~(\ref{Ofss}), see
\cite{SupplementaryMaterial}.  Note that the approach to the OS curves
requires the necessary condition $L\gg \xi_\pm$ where $\xi_\pm$ are
the correlation lengths of the pure phases at $T_c^\pm$ ($\xi_-\approx
\xi_+= 2.695$ for $q=20$~\cite{results-q20}).

\begin{figure}[tbp]
\includegraphics*[scale=\graphicscale]{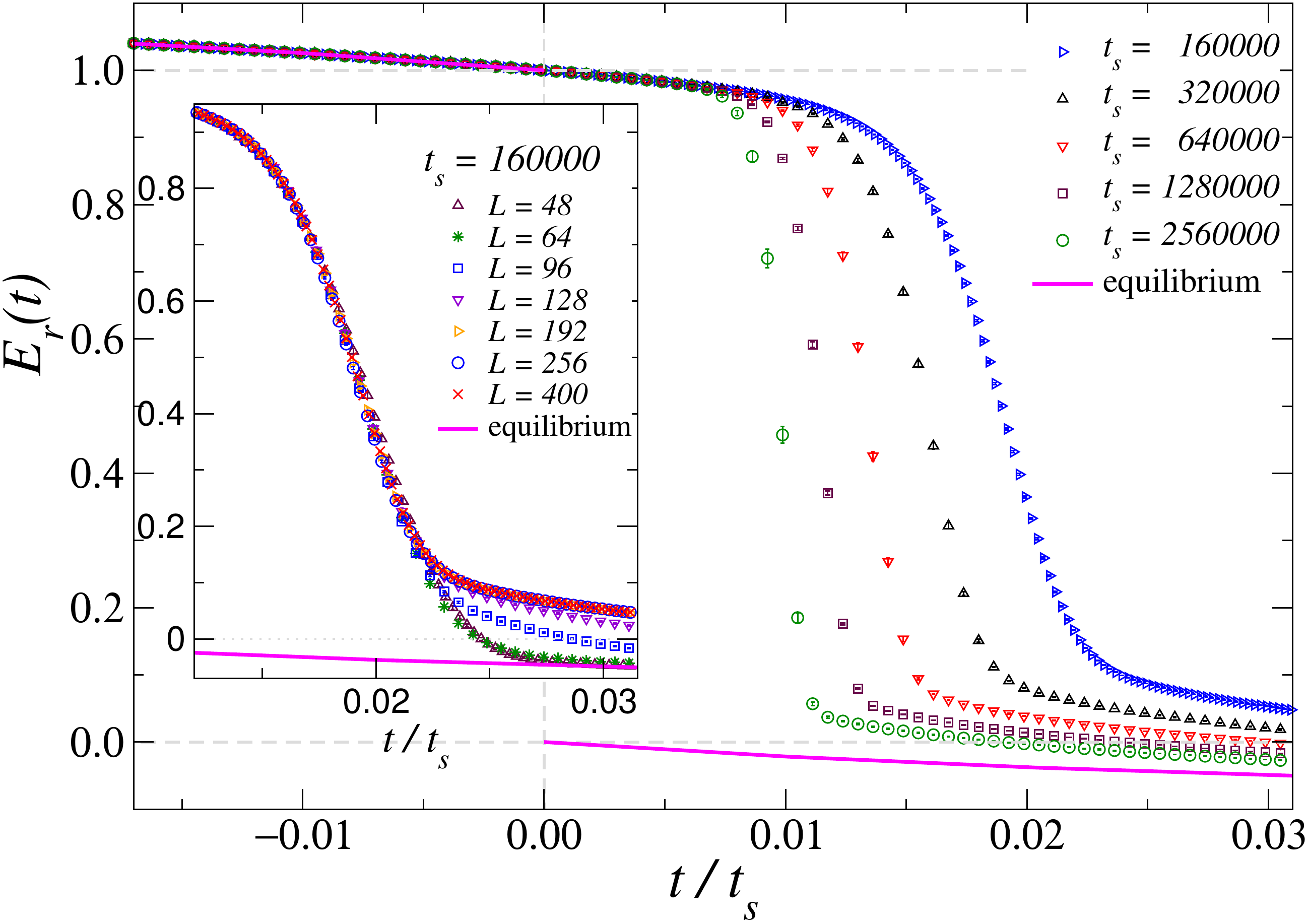}
\caption{(Color online) MC data of $E_r(t)$ for $q=20$ for some values
  of $t_s$, in the $L\to\infty$ limit. The large-$L$ convergence
  (within errors) is checked by increasing $L$ at fixed $t_s$,
  analogously to the case shown in the inset.  The full lines show the
  equilibrium energy density at $\beta = \beta_c(1 + t/t_s)$.  }
\label{infvlim}
\end{figure}

We now show that an interesting off-equilibrium behavior develops in
the infinite-volume limit, corresponding to $s_1\to 0$.  As shown by
Fig.~\ref{infvlim}, data at fixed $t_s$ have a well-defined large-$L$
limit, see the inset of Fig.~\ref{infvlim}.  This is rapidly
approached for small values of $\delta(t)\equiv t/t_s$, e.g.,
$\delta(t)\lesssim 0.02$ at $t_s\approx 10^5$, while significantly
larger lattices are required for larger $\delta(t)$.  The energy
density does not converge to its equilibrium value as $L\to \infty$,
due to the fact that the system settles in a metastable state with
large coexisting droplets of different colors.

\begin{figure}[tbp]
\includegraphics*[scale=\graphicscale]{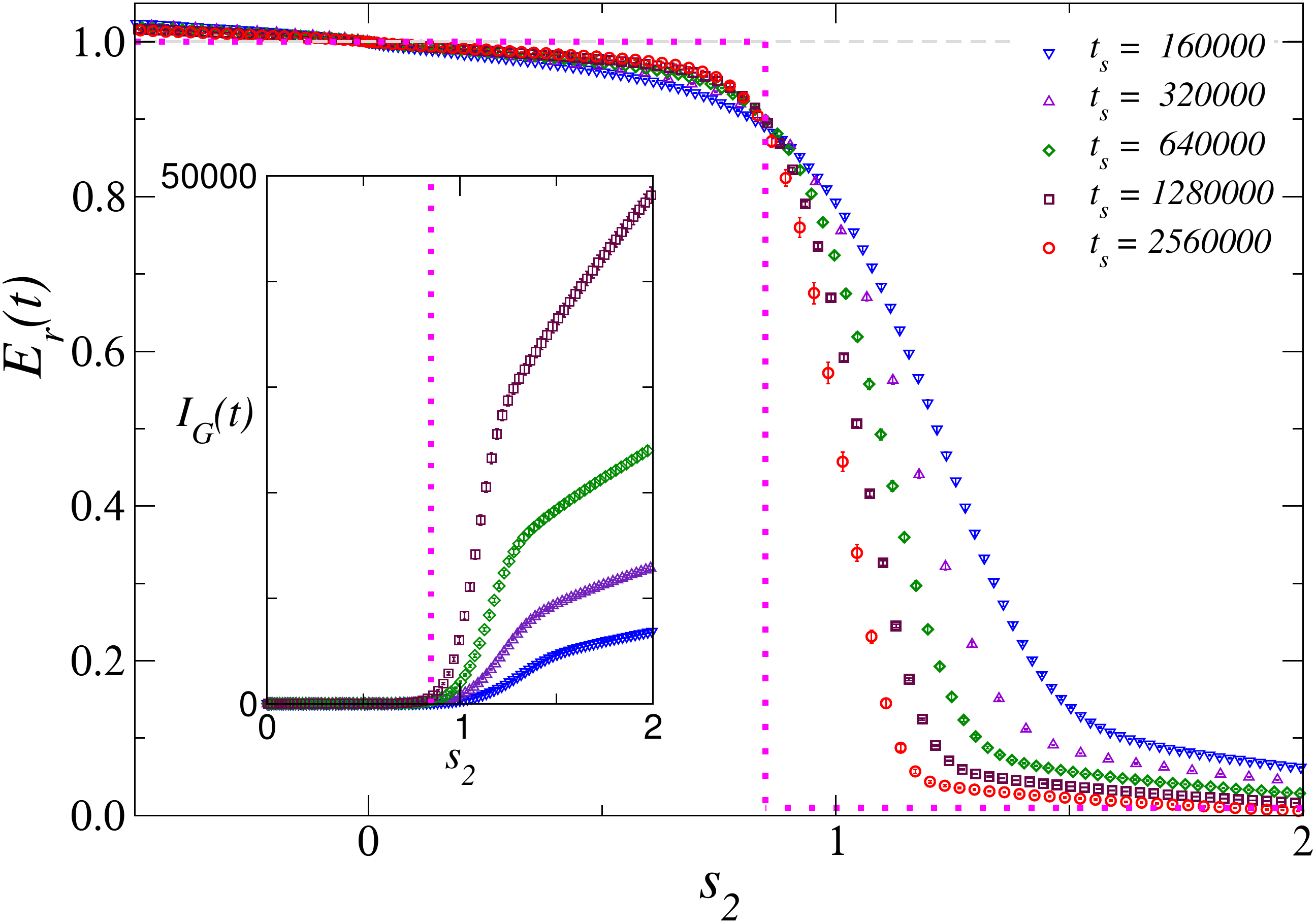}
\caption{(Color online) The infinite-volume $E_r$ and $I_G$ (inset)
  for $q=20$ versus $s_2$.  The dotted lines show the conjectured
  singular large-$t_s$ limit, see text.  
}
\label{infvo}
\end{figure}

The infinite-volume energy density (see Fig.~\ref{infvlim}) takes the
equilibrium high-$T$ value $E_r(t=0)=1$ at $t=0$ for any $t_s$, then
shows a sharp decrease at a point $\delta^*(t_s)$, which decreases
with increasing $t_s$.  The system develops a nontrivial OS behavior
close to $\delta^*(t_s)$.  For large $L$ the system behaves as a gas
of droplets of size $R$ (evidence for this behavior is provided in
\cite{SupplementaryMaterial}).  The relevant scaling variables are
expected to be analogous to $r_1$ and $r_2$, cf.~Eq.~(\ref{s12}), with
$R$ replacing the size $L$.  The relevant time scale is that of the
formation of droplets of size $R$. As the time $\tau_d$ to create a
droplet of size $R$ increases exponentially with $R$, $\ln \tau_d \sim
R$, we expect $R\sim \ln t$.  Thus, the analogue of the scaling
variable $r_1$ becomes ($t > 0$)
\begin{equation}
s_2 = (t /t_s) \ln^2 t.
\end{equation}
In Fig.~\ref{infvo} we report the infinite-volume energy density and
$I_G$ for $q=20$ versus $s_2$.  We note a crossing point of the energy
curves for different values of $t_s$ at approximately $s_2^*\approx
0.85$ with $E_r^*\approx 0.89$. At the same value of $s_2$, $I_G$
shows a sharp change of behavior. These results suggest that, in the
limit $t_s\to \infty$, the OS functions develop a singular behavior
for $s_2 = s_2^*$.  In particular, the infinite-volume energy density
takes the high-$T$ value ${\cal E}_\infty(s_2)=e_+ = 1$ for $s_2 <
s_2^*$, while we expect ${\cal E}_\infty(s_2)=e_- \ll 1$ for $s_2 >
s_2^*$.  Note that, for large $t_s$, $(t \ln^2 t)/t_s = s_2^*$ implies
$t/t_s \approx s_2^*/(\ln t_s)^2$, so that the value $\beta_d$ of
$\beta$ at which the sharp change occurs converges to $\beta_c$ as
$t_s$ increases.

\begin{figure}[tbp]
\includegraphics*[scale=\graphicscale]{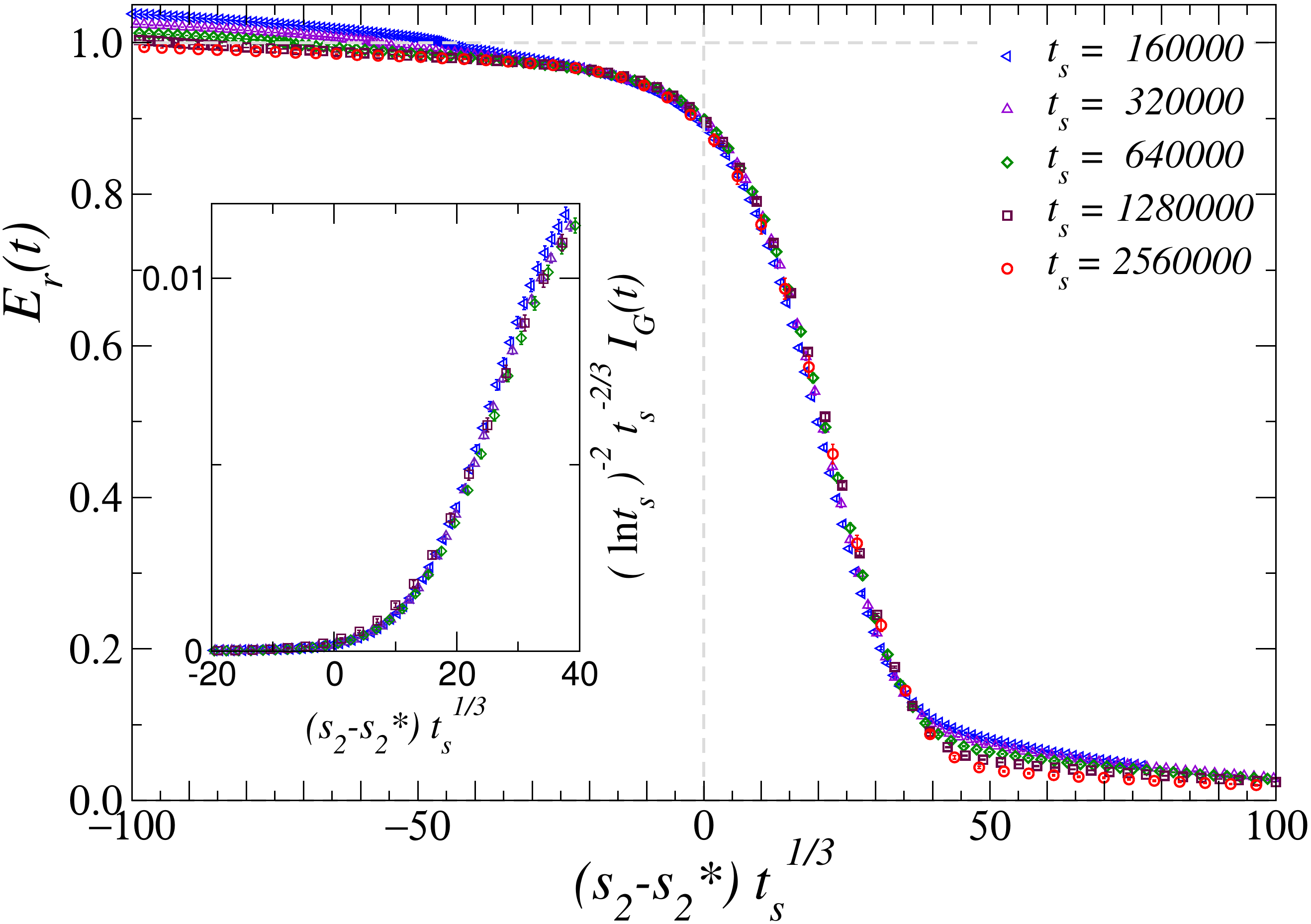}
\caption{(Color online) Scaling of the infinite-volume $E_r$ and $I_G$
  (inset) around $s_2^*$. }
\label{infvostar}
\end{figure}

The behavior around $s_2^*$ turns out to be described by an additional
scaling Ansatz. As shown in Fig.~\ref{infvostar}, the energy density
${\cal E}_\infty(s_2,t_s)\equiv E_r(t,t_s,L\to\infty)$ scales as
\begin{equation}
{\cal E}_\infty(s_2,t_s) \approx f_e(\tilde{s}_2),
\qquad \tilde{s}_2 = (s_2-s_2^*) t_s^\theta
\label{estar}
\end{equation}
with~\cite{footnotebias} $\theta = 1/3$.  
We stress that scaling is only observed when using the variable $s_2$.
The estimate $\theta=1/3$ is reasonably accurate (10\% accuracy).
Also $I_G(s_2,t_s)$ shows a scaling behavior, provided we multiply it
by an additional power of $t_s$. Phenomenologically we observe $I_G(t)
\approx (\ln t_s)^2 t_s^{2/3} f_G (\tilde{s}_2)$, see the inset of
Fig.~\ref{infvostar} (the exponents of $t_s$ and $\ln t_s$ in the
prefactor are an educated guess).  A similar analysis can be performed
for $q = 10$, see \cite{SupplementaryMaterial}.  The estimate of
$s_2^*$ changes ($s_2^* \approx 0.2$ for $q=10$), but all other
conclusions hold.  In particular, MC data are again consistent with
$\theta=1/3$.  This singular behavior resembles that at the mean-field
spinodal point~\cite{Binder-87}, or more generally the power-law
scaling at equilibrium continuous transitions.  However, here the
location $\beta_d$ of the dynamic transition converges to $\beta_c$ as
$t_s\to \infty$: $\beta_d -\beta_c \sim (\ln t_s)^{-2} \to
0$~\cite{footnotepotts}.

\begin{figure}[tbp]
\includegraphics*[scale=\graphicscale,angle=-90]{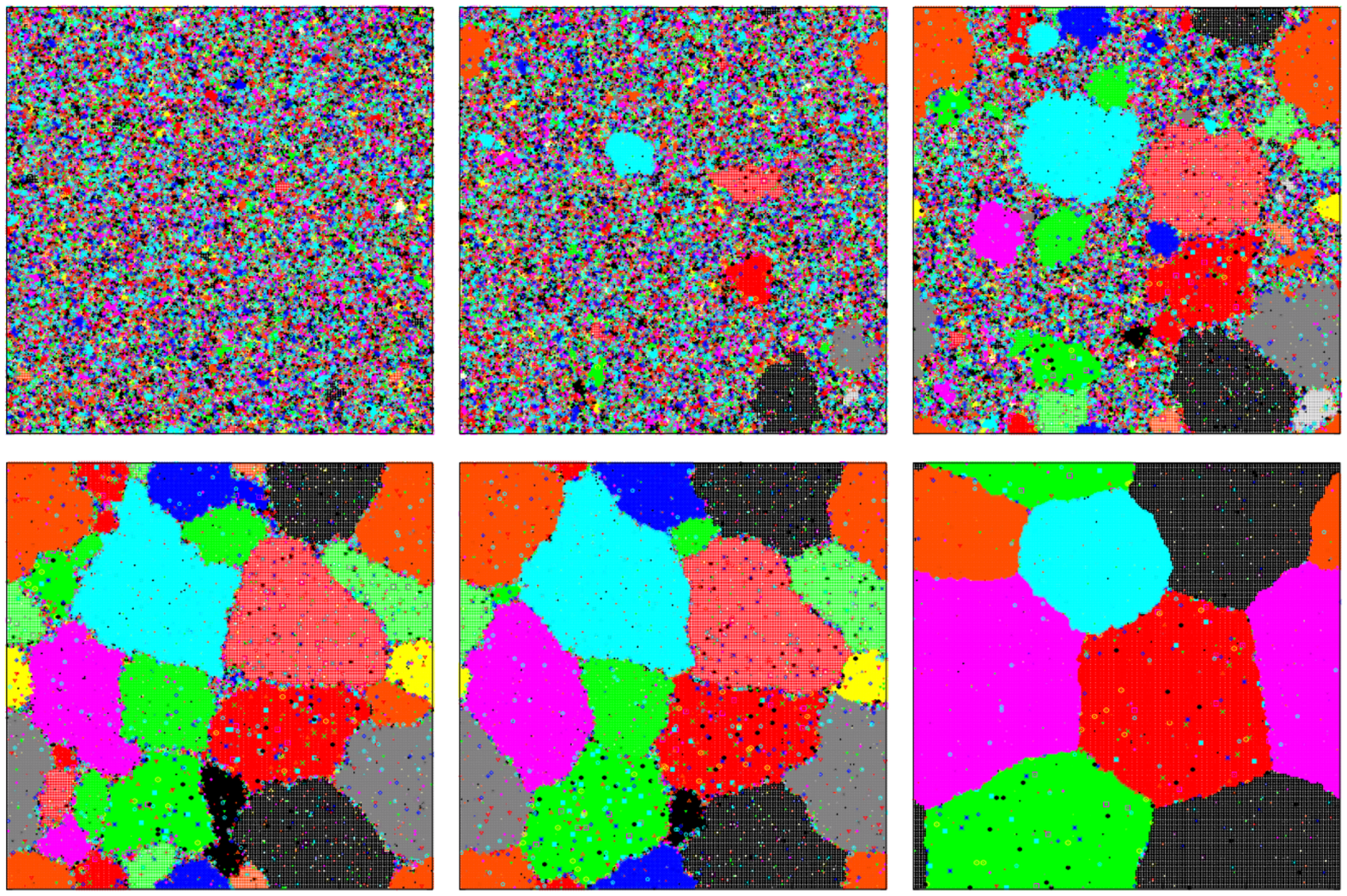}
\caption{(Color online) Snapshots of a system of size $L=512$ for
  $\tilde{s}_2=(s_2-s_2^*)t_s^{1/3}= -20,\,0,\,20,\,40,\,100,\,1500$,
  (from left to right, top to bottom). Here $q=20$ and $t_s=640000$.
  We use different colors for each value of $s_x$.  The range of
  $\tilde{s}_2$ covers the region where $E_r$ significantly changes,
  see Fig.~\ref{infvostar}.}
\label{droplets}
\end{figure}

\begin{figure}[tbp]
\includegraphics*[scale=\graphicscale,angle=0]{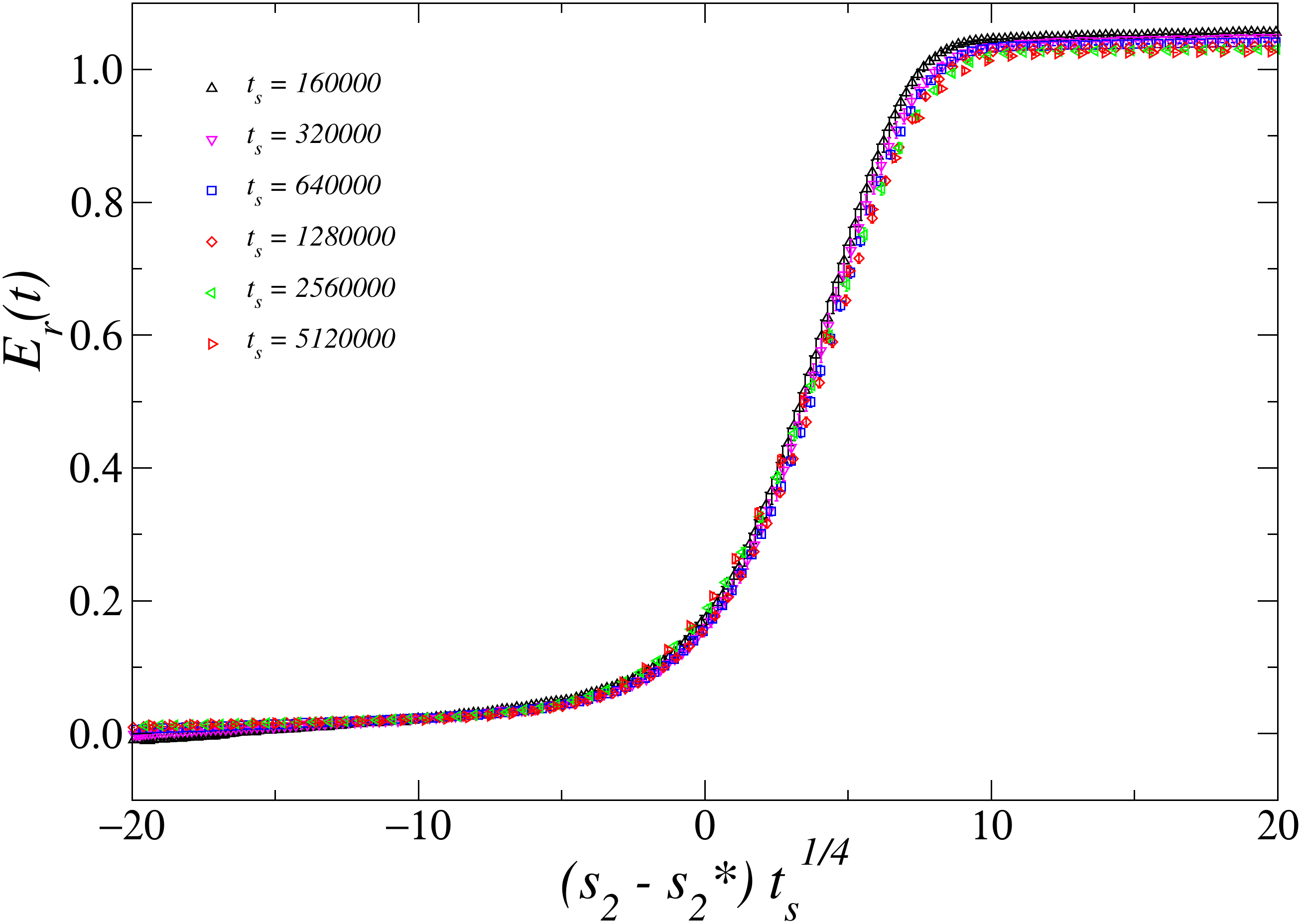}
\caption{(Color online) Scaling of the infinite-volume $E_r$ for the
  reversed protocol, in which the dynamics starts from an ordered
  configuration at $\beta_i > \beta_c$ and $\beta(t) = \beta_c (1 -
  t/t_s)$.  The optimal collapse is obtained for $s_2^* \approx 0.85$
  and $\theta \approx 1/4$ [the accuracy of the data does not exclude
    the value $\theta\approx 1/3$ obtained for protocol
    (\ref{deltat})], see \cite{SupplementaryMaterial}.}
\label{reverse-protocol}
\end{figure}

To understand the behavior of the system for $s_2\approx s_2^*$, one
may consider the evolution of the size of the clusters formed by spins
of the same color.  A typical case is reported in Fig.~\ref{droplets}.
For $s_2 \lesssim s_2^*$, the system is disordered and all clusters
are small: their typical size $\ell_d$ satisfies $\ell_d \lesssim
\xi_+$, where $\xi_+$ is the correlation length of the pure disordered
phase~\cite{results-q20}.  For $s_2\approx s_2^*$ clusters start
growing. There is a short coarsening
interval~\cite{Bray-94,CEGS-12,Biroli-15}, in which $E_r$ decreases
almost linearly. Then, the system settles in a metastable state
characterized by many coexisting large clusters.  Details are reported
in \cite{SupplementaryMaterial}.

One may also consider the dynamics induced by the reverse linear
protocol across $\beta_c$, starting from an ordered configuration
(equal spins) at $\beta_i>\beta_c$ and decreasing $\beta$ across the
FOT, $\beta(t) = \beta_c (1 - t/t_s)$, up to $\beta_f<\beta_c$.  In
this case the $q$-permutation symmetry is broken by the initial
condition.  MC results show an analogous OS behavior.  In the
infinite-volume limit the system persists in the ordered state up to a
temperature $T(t_s) > T_c$, where it transits to the disordered
state. For $T\approx T(t_s)$ one observes analogous OS laws.
Eq.~(\ref{estar}) holds with a corresponding $s^*_2$ and exponent
$\theta$, which may differ from those of protocol (\ref{deltat}), see
Fig.~\ref{reverse-protocol} and \cite{SupplementaryMaterial}.

The OS theory can be applied to hysteresis phenomena that occur when
$T$ is first decreased below $T_c$ and then increased above $T_c$.
We mention that external periodically-varying fields have been already
considered at magnetic FOTs, where they give rise to a singular
dynamic behavior of the magnetization hysteresis~\cite{CA-99,SRN-98}.

In conclusion, we have developed an OS theory to describe the
off-equilibrium behavior of statistical systems when their temperature
is slowly varied across a thermal FOT.  We consider the linear
protocol (\ref{deltat}) and the reversed one~\cite{footnoteabslin}.
Our numerical study of the Potts model confirms the general OS theory.
In particular, in the infinite-volume limit it shows two dynamic
regimes, separated by a spinodal-like transition point where the OS
functions are singular.  Such a transition occurs at a time $t_d>0$
scaling as $t_d\sim t_s (\ln t_s)^{-2}$ in the large-$t_s$ limit.
Therefore, a spinodal-like behavior emerges dynamically in short-range
models, without assuming long-range interactions as in the mean-field
theory~\cite{Binder-87}.  The OS behavior arises from the interplay
between the exponentially large tunneling times at $T_c$ and the
droplet formation.  We expect that analogous dynamic scaling behaviors
emerge at any thermal FOT characterized by these two features.
Further investigations are needed to clarify its degree of
universality and to develop a theory which is able to predict the
exponent $\theta$ entering the OS laws, such as Eq.~(\ref{estar}).

Our results provide an effective framework to interpret experimental
data in many physical contexts, when thermal FOTs are crossed by
slowly varying $T$.  For example, we mention the formation of the
quark-gluon plasma in heavy-ion collisions~\cite{qgp-ref}, whose
intrinsic space-time inhomogeneities complicate the study of the
hadronic phase diagram, and, in particular, the expected thermal FOT
line at nonzero baryon chemical potential~\cite{RW-00}.  Another issue
concerns the universe evolution.  Kibble~\cite{Kibble-76} made the
first analysis of the behavior of a system going across a continuous
transition, to study the defect production during the universe
expansion.  Analogous studies at FOTs may shed some light on the
behavior of an expanding and cooling universe going across a
FOT~\cite{BVS-06}.


\begin{thebibliography}{99}
\bibitem{footnotepre} This is obtained by an educated guess after
  looking at the data. Unbiased analyses confirm it with 20\%
  accuracy. We note that the precision on the exponent $\alpha$ of the
  prefactor of $\tau(L)=L^\alpha e^{\sigma L}$ is not crucial for the
  scaling behavior, which is essentially controlled by the dominant
  exponential dependence.
\bibitem{footnotebias} We are biased by the idea that simple integer
  or fractional exponents enter the various off-equilibrium scaling
  ansatzs at FOTs, likely arising from geometric reasons.
\bibitem{footnoteabslin} Linear time variations are quite general
  since any generic time dependence can be linearly approximated
  around $T_c$.  If the linear term is present, higher-order
  contributions are suppressed in the OS limit (this can be shown by
  scaling arguments); if it is absent, the OS behavior changes, but it
  can still be derived by a generalization of the scaling arguments.
\end{thebibliography}
\end{document}